\documentclass[12pt]{iopart}
\usepackage{graphicx, bm, color, cite}
\usepackage{iopams}
\usepackage[nolists]{endfloat}

\newcommand{\transpose}{^\mathrm{T}}
\newcommand{\expm}[1]{\exp\!\big(#1\big)}
\newcommand{\expt}[1]{\langle #1 \rangle}

\newcommand{\be}{\begin{eqnarray}}
\newcommand{\ee}{\end{eqnarray}}

\newcommand{\beq}{\begin{equation}}
\newcommand{\eeq}{\end{equation}}

\newcommand{\ta}{\tau_1}
\newcommand{\tb}{\tau_2}

\begin{document}

\title{Gene-history correlation and population structure}

\author{A. Eriksson\dag\ and B. Mehlig\ddag}
 \address{\dag\ Dept. of Physical Resource Theory, Chalmers and G\"oteborg
 University, Sweden}
 \address{\ddag\ Dept. of Theoretical Physics, G\"oteborg
 University and Chalmers, Sweden}

\begin{abstract}
Correlation of gene histories in the human genome determines the
patterns of genetic variation ({\em haplotype structure}) and is
crucial to understanding genetic factors in common diseases. We
derive closed analytical expressions for the correlation of gene
histories in established demographic models for genetic evolution
and show how to extend the analysis to more realistic (but more
complicated) models of demographic structure. We identify two
contributions to the correlation of gene histories in divergent
populations: linkage disequilibrium, and differences in the
demographic history of individuals in the sample. These two
factors contribute to correlations at different length scales: the
former at small, and the latter at large scales. We show that
recent mixing events in divergent populations limit the range of
correlations and compare our findings to empirical results on the
correlation of gene histories in the human genome.
\end{abstract}

\submitto{Physical Biology}

\maketitle

\clearpage \newpage %
\section{Introduction}
\label{sec:introduction}

Populations are shaped by demographic, historical and
social factors, determining gene histories in characteristic ways.
Empirical data on genetic variation are now routinely interpreted
using well-established gene-genealogical models
\cite{hudson90,nordborg_tavare02,reich_etal02,hapmap_group03} of
the population in question. Local properties of genetic variation
(pertaining to {\em loci}, short stretches of a chromosome) in
such models are very well understood, by means of models of
bottlenecks, population expansion \cite{tajima87a, tajima87b,
slatkin_hudson91, sano_etal04}, and migration \cite{wakeley96,
teshima_tajima03, stumph_goldstein03}.
By contrast, very little is know about global patterns
\cite{patil_etal01}.
Global correlation and variation of patterns appear to be the key
to understanding the genetic factors contributing to common
diseases: there is now a wealth of empirical information on the
variation of genetic material in the human genome
\cite{snp_group01}. Many common diseases (such as cancer, obesity,
cardiovascular disorder and diabetes) are caused by combinations
of genetic and environmental factors \cite{hapmap_group03}. In
some cases a common variant of a single gene is responsible for
specific syndromes. In more complex diseases, however, it may not
be possible to link a disease to a single genetic factor. It is
thus necessary to understand genome-wide association of genetic
factors.

\begin{figure}
   \centerline{\includegraphics{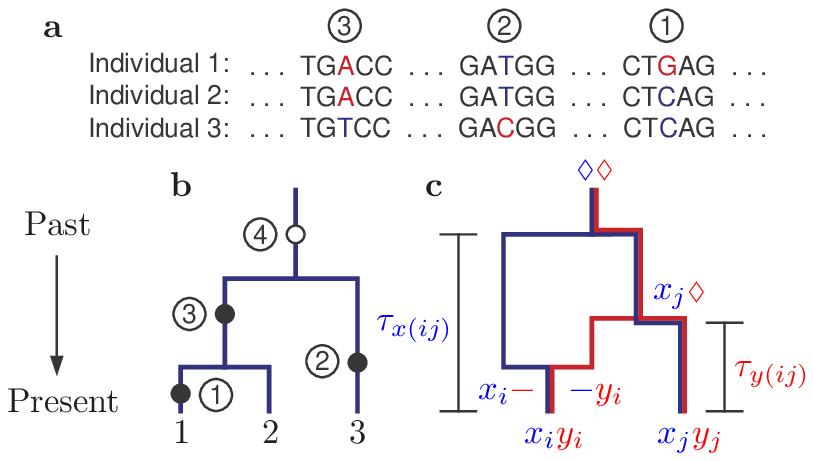}}
\caption{\label{fig:samplegenealogy}
Gene history and polymorphic sites. \textbf{a} In DNA, genetic
information is encoded by base-pairs of the four nucleic acids
adenine ({\tt A}), thymine ({\tt T}), guanine ({\tt G}), and
cytosine ({\tt C}). In a sample of three individuals, we show
three polymorphic sites, with two nucleotides around each
polymorphism. \textbf{b} The most common variation is a difference
at a single position (SNP), caused by a mutation at the position
in an individual in the history of the population, where e.g. a
fraction of the population has the nucleotide {\tt T} at the site,
and the rest has the nucleotide {\tt A}. The three mutations in
panel \textbf{a} are shown as filled circles. Mutation 4 does not
cause a polymorphism in the sample, since all individuals in the
sample inherits the mutation from the common ancestor. Given
$\tau$ (the number of generations since the most recent common
ancestor) of a stretch of $L$ nucleotides, the number of
differences between two individuals is assumed to be Poisson
distributed with expected value $2 \mu L \tau$, where $\mu$ is the
mutation rate per site per generation \cite{hudson90}. \textbf{c}
In recombination, part of a \emph{gamete} (one of the two copies
of a chromosome) is inherited from one parent and the rest from
the other parent. We show a sample gene history with one
recombination event, for two loci ($x$ and $y$) in two gametes
$i$ and $j$. The time axis is the same
as in panel \textbf{b}. The ancestral history for
loci $x$ and $y$ are shown in blue and red, respectively. The
times until the most recent common ancestor are $\tau_{x(ij)}$ and
$\tau_{y(ij)}$ for loci $x$ and $y$, respectively. In the absence of
recombination, two loci on the same gamete share the same genetic
history, and have the same time to the most recent common
ancestor, $\tau_{x(ij)} = \tau_{y(ij)}$, causing \emph{linkage
disequilibrium}. If a recombination event occurs in the genetic
history of a sample, it may lead to a decorrelation of $\tau_{x(ij)}$
and $\tau_{y(ij)}$. 
$x_i$ represents the genetic material at locus $x$ of
chromosome $i$. Dashes correspond to genetic material not in the
history of the sample, and the diamonds to common ancestral
material.
}
\end{figure}

Mutations and linkage disequilibrium (explained and illustrated in
figure~\ref{fig:samplegenealogy}) determine the genetic history of
a population, which in turn shapes the patterns of genetic
variation of interest in gene association studies
\cite{patil_etal01,hapmap_group03}.
The question is: how strongly are the patterns at two different
loci correlated?
Reich \etal \cite{reich_etal02} estimate the empirical association
of polymorphism rates, as a function of the physical distance
between the loci on the same chromosome, from human population
data (compensating for variations in the mutation rate along the
chromosome by comparing to the population data from the great
apes). Assuming a neutral model with uniform mutation rate, the
covariance of polymorphism rates is given by the covariance of the
times to the most recent common ancestor of the two loci (c.f.
figure \ref{fig:samplegenealogy}c).
Kaplan and Hudson \cite{kaplan_hudson85} (see also
\cite{hudson83}) analysed the association of polymorphism rates for
short loci, within the standard unstructured neutral model. This
was further developed by Pluzhnikov and Donelly
\cite{pluzhnikov_donelly96}, who analysed optimal sample sizes for
surveying genetic diversity.
Hudson \cite{hudson01} and McVean \etal \cite{mcvean_etal02}
estimate the recombination rate likelihood from two-locus sample
statistics, based on simulations. Recombination rate likelihoods,
conditional on more than two sites, have also been estimated using
Monte-Carlo methods
\cite{griffiths_marjoram96,kuhner_etal00,nielsen00}. Although
statistically powerful, these methods are computationally very
demanding.
Linkage disequilibrium is often assessed through summary
statistics such as $r^2$ \cite{hill_robertson68} or $D'$
\cite{tajima87a}. McVean \cite{ mcvean02} introduced an
approximation $\sigma^2_d$ of the expected value of $r^2$, and
showed that the approximation is accurate, in the absence of
demographic structure, if the expectations are taken conditional
on intermediate allelic frequencies.

\begin{figure}
   \centerline{\includegraphics{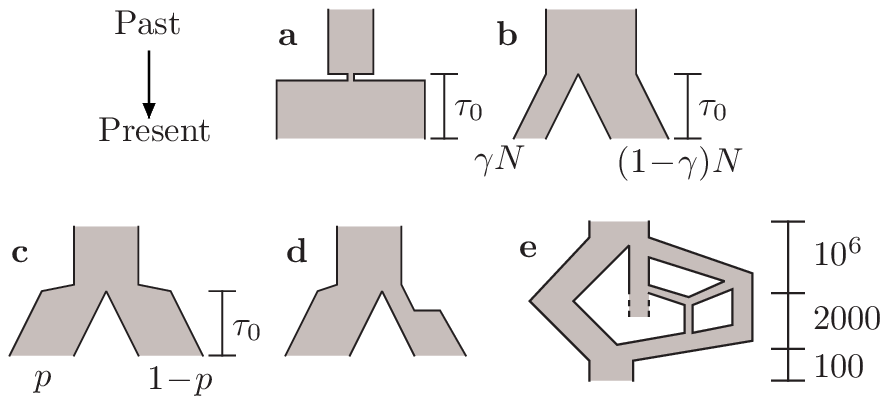}}
\caption{\label{fig:pop struct models}
Models illustrating demographic history, i.e. changes in
population size and structure. \textbf{a} Population bottleneck.
\textbf{b},\textbf{c} Models of population structure and
expansion. \textbf{d} A more general model of demographic
structure. \textbf{e} Demographic structure determining genetic
variation in the laboratory-mouse genome \cite{wade_etal02} (time
here is measured in years).
}
\end{figure}

In this paper, we derive analytical expressions for the
correlation of genetic histories in established models of
demographic history (see figure~\ref{fig:pop struct models}a--c)
in the limit of negligible selection.
For several reasons these results are of interest.
First, as explained in the following, they enable us
to gain a qualitative understanding of the relative importance
of different biological factors determining the empirically
observed patterns of linkage disequilibrium. Second,
the analytical results summarised in this article
can be easily generalised as explained below
(see figure~\ref{fig:pop struct models}d,e).
Third, our analytical expressions for the decorrelation
of gene histories allow for studying the implications
of variations of the recombination rate along the chromosomes
\cite{kong_etal02,eriksson_mehlig04}.
The remainder of this paper is organised into five  parts. We
begin by discussing gene-history correlations and linkage
disequilibrium in section \ref{sec:gene-history correlations}
(see also figure~\ref{fig:samplegenealogy}). In section \ref{sec:methods}
we describe our method. We summarise our results in section
\ref{sec:results} and discuss their implications in section
\ref{sec:discussion}. In section \ref{sec:conclusions} we draw
conclusions. Two appendices summarise details of our calculations.


\clearpage \newpage %
\section{Gene-history correlations, linkage disequilibrium, and
patterns of genetic variation}
\label{sec:gene-history correlations}

Genetic variation is caused by multiple factors. Together,
mutations and recombination (figure~\ref{fig:samplegenealogy}) are
the most important determinants of the large-scale haplotype
structure in the human genome \cite{reich_etal02, patil_etal01,
hapmap_group03}. The genetic history of nearby sites is closely
related, while distant sites may become unrelated only a few
generations in the past. 

Correlation of gene histories determines the degree of association
between patterns of genetic variation at different loci.
An example is the correlation of the counts of
single-nucleotide polymorphisms (SNPs) at different loci:
let $S_{x(ij)}$ be the number of SNPs
at locus $x$ between a pair of chromosomes $i$ and $j$.
Further, let $\tau_{x(ij)}$ denote
the time to the most recent common ancestor of a locus at position
$x$ on chromosomes $i$ and $j$, and define $\tau_{y(ij)}$
correspondingly for the locus at position $y$.
Then the sample covariance of the number of SNPs
in non-overlapping loci $x$ and $y$ is
related to the covariance of times $\tau_{x(ij)}$ and $\tau_{y(ij)}$ as
follows
\begin{equation}\label{eq:cov S_a S_b}
   \mathrm{cov}[S_{x(ij)},S_{y(ij)}] \approx (2 \mu L)^2 \, \mathrm{cov}[\tau_{x(ij)},\tau_{y(ij)}]\,.
\end{equation}
Here  $L$ is the size of the loci, assuming variations in the
mutation rate $\mu$ along the chromosome are negligible. For
(\ref{eq:cov S_a S_b}) to hold, $L$ must be small enough that the
sites within each locus have a high degree of linkage (in humans,
$L$ must be of the order of or smaller than a few hundred
base-pairs).

Associations between SNPs in the genetic mosaic
allows for efficient mapping of genes. Suitably
chosen, a relatively small set of SNPs can capture most of the
common patterns of variation in the genome \cite{hapmap_group03}.

The decay of the covariance $\mbox{cov}[\tau_{x(ij)},\tau_{y(ij)}]$ as a
function of $|x-y|$ measures linkage disequilibrium. 
In the remainder of this section we briefly comment on other
common measures of linkage disequilibrium. Global association
between patterns of diversity, quantified by the extent of linkage
disequilibrium is often measured by Tajima's $D'$ \cite{tajima87a} or
alternatively by
\beq
   r^2 = \frac{D^2}{f_{A(x)} (1 - f_{A(x)}) f_{B(y)} (1 - f_{B(y)})},
\eeq
where $D = f_{A(x)B(y)} - f_{A(x)} f_{B(y)}$, $A(x)$ and $B(y)$ are 
the allelic types at the loci $x$ and $y$, respectively, and
$f_{A(x)B(y)}$ is frequency of alleles $A(x)$ and $B(y)$ on the
same chromosome in the sample \cite{tajima87a}. McVean
\cite{mcvean02} introduced an approximation to the expected value
of $r^2$, called $\sigma^2_d$, which makes the connection to the
correlation of gene history explicit. With the notation $E_{ij,kl}
= \expt{ \tau_{x(ij)} \tau_{y(kl)}}$,
\beq\label{eq:sigma2 def}
   \sigma^2_d =
   \frac{ (n^2 - 2n + 2)E_{ij,ij} - 2(n-2)^2 E_{ij,ik} + (n-2)(n-3) E_{ij,kl} }
        { 2 E_{ij,ij} + 4(n-2) E_{ij,ik} + (n - 2)(n - 3) E_{ij,kl}} \, .
\eeq
The factors $E_{ij,ij}$ and $E_{ij,ik}$ are defined analogously.
For unstructured populations, $\sigma^2_d$ and the expected value
of $r^2$ are approximately equal under the neutral dynamics, if
the expectation is conditioned on intermediate allelic frequencies
\cite{mcvean02}.

\clearpage \newpage %
\section{Methods}
\label{sec:methods}

In the following we analyse how correlation of gene histories
depends on demographical factors. In a large, unstructured population
with constant population size, and when selection is negligible,
the ancestral history of a locus may be modeled as a Markov
process \cite{griffiths81, hudson_kaplan85, nordborg_tavare02},
where the states of the process correspond to different
configurations of ancestral DNA through the history of the sample.

We trace the ancestral history of two loci (at positions $x$ and
$y$) in $n$ individuals, from the present
back in time until the most recent common ancestor has been found
for all loci. When the population size $N$ is large, the genealogical
process may be approximated by the so-called coalescent process \cite{hudson90}:
recombination is modeled as a Poisson
process with rate $r$ per generation per chromosome: for any given
chromosome, with probability $r$ (also known as the recombination
fraction) the loci stem from different parents. The
probability that one pair of individuals has a common ancestor in
the preceding generation, and the probability that an individual
inherits genetic material from both parents, are expanded in
$N^{-1}$ to the first order. Time is measured in units of $2N$
generations. In the limit of large $N$, the time to the next event
is approximately exponentially distributed \cite{hudson90}.

\begin{figure}
\centerline{
\begin{tabular}{@{}ll@{}}
   \includegraphics{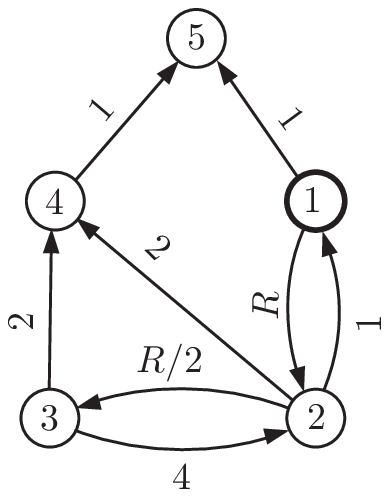}
   &
   \raisebox{2.5cm}{
   \begin{tabular}{cl}
      \br
      \small State $i\ $ & \small Population \\
      \mr
      \small\raisebox{1.5ex}{$1$} & \small\shortstack{$x_iy_i$,\,$x_jy_j$\\
                    $x_iy_j$,\,$x_jy_i$} \\
      \mr
      \small\raisebox{4ex}{$2$} & \small\shortstack{
                    $x_i-$,\,$-y_i$,\,$x_jy_j$\\
                    $x_iy_i$,\,$x_j-$,\,$-y_j$\\
                    $x_i-$,\,$-y_j$,\,$x_jy_i$\\
                    $x_iy_j$,\,$x_j-$,\,$-y_i$
                 }\\
      \mr
      \small$3$ & \small$x_i-$,\,$-y_i$,\,$x_j-$,\,$-y_j$ \\
      \mr
      \small\raisebox{1ex}{$4$} & \small\shortstack{$x_i\scriptstyle\lozenge$,\,$x_j\scriptstyle\lozenge$\\
                    ${\scriptstyle\lozenge}y_i$,\,${\scriptstyle\lozenge}y_j$}\\
      \mr
      \small$5$ & \small$\scriptstyle\lozenge\lozenge$ \\
      \br
   \end{tabular}
   }
\end{tabular}
}
\caption{\label{fig:markovgraph}
A graph representation of the coalescent process for two loci ($x$
and $y$) and two chromosomes ($i$ and $j$). The transition rates
(measured in units of $2N$ generations) between the different
groups of states, corresponding to the table, are printed along
the arrows ($R = 4Nr$). The process starts in state $1$ and
ends in state $5$, the only absorbing state. If the path goes from
state $1$ to state $5$ we have linkage, but if the system enters
state $4$ linkage is broken. 
Same notation as in figure~1.
}
\end{figure}

By explicitly taking into account the symmetries of the state
space of the coalescent for two individuals, we obtain a compact
representation of the Markov process
(figure~\ref{fig:markovgraph}) which allows us to derive and
understand gene-history correlations in the models mentioned
in the introduction.

We illustrate our approach by re-deriving Hudson's result for the
correlation of gene histories in the unstructured, constant
population-size coalescent model \cite{hudson83}. Consider a
sample of two individuals. Figure~\ref{fig:markovgraph} shows a
representation of the coalescent for this case. Each node in the
graph corresponds to a configuration of ancestral DNA (listed in
the table in figure~\ref{fig:markovgraph}). Due to the symmetries
of the coalescent, many different configurations may be mapped
onto the same node.

The time evolution of the probability distribution $P_i(t)$ over the states
$i$ is given by the master equation
\begin{equation}
   \partial_t P_i(t) = \sum_j w_{j \rightarrow i} P_j(t) - \sum_j  w_{i \rightarrow j} P_i(t)\,,
\end{equation}
where $w_{i \rightarrow j}$ is the transition rate from state $i$
to state $j$, given in figure~\ref{fig:markovgraph}. As above, time is
measured in units of $2N$ generations. The process is started in
state $1$, and proceeds until it comes to state $5$. We find that
$\langle\tau_{x(ij)}\tau_{y(ij)}\rangle$ is given by the exit rates to state
$5$, via states $1$ and $4$. Let $\ta$ be the first time at which a locus
coalesces, and $\tb$ be the time when both loci have coalesced.
Since $\tau_{x(ij)}\tau_{y(ij)} = \ta\tb$ we obtain
\begin{equation}
\label{eq:corr}
   \left<\tau_{x(ij)}\tau_{y(ij)}\right> =
      \int_0^\infty \Big[
      {\bm u}_1\transpose\tau_1^2
      + {\bm u}_2\transpose\!
      \int_{\tau_1}^\infty \tau_1 \tau_2\,
      {\rm e}^{\tau_1\!-\!\tau_2} \,\rmd\tau_2
      \Big] {\rm e}^{{\bf M} \tau_1} \,{\bm v} \,\rmd\tau_1 \,,
\end{equation}
where ${\bm v} = {\bm u}_1 = (1,0,0)\transpose$, ${\bm u}_2 =
(0,2,2)\transpose$ and $\mathbf{M}$ is a three-by-three matrix
defined by $\mathbf{M}_{ij} = w_{j \rightarrow i}$ for $i,j = 1,
\dots, 3$ and $i \ne j$, and $M_{ii} = - \sum_{j=1}^{3} w_{i
\rightarrow j}$. Evaluating (\ref{eq:corr}) we obtain the
well-known result \cite{hudson_kaplan85,hudson83}
\begin{equation}\label{eq:rho_no_pop_struct}
   \rho(\tau_{x(ij)},\tau_{y(ij)}) \equiv
   \frac{\left<\tau_{x(ij)}\tau_{y(ij)}\right> - \left<\tau\right>^2}{ \left<\tau^2\right>
   -  \left<\tau\right>^2} =  \frac{R + 18}{R^2 + 13 R + 18} \,,
\end{equation}
where $R = 4Nr$. In order to calculate $\sigma^2_d$ for the
unstructured model, we obtain $\expt{\tau_{x(ij)}\tau_{y(ik)}}$
and $\expt{\tau_{x(ij)}\tau_{y(kl)}}$ from (\ref{eq:corr}) with
${\bm v} = (0,1,0)\transpose$ and ${\bm v} = (0,0,1)\transpose$,
respectively. Inserting these into eq.~(\ref{eq:sigma2 def}), we recover
the result of McVean \cite{mcvean02}:
\beq
   \sigma^2_d =
 \frac{2\,( 6 + R )  + n\,( 10 + 11 R + R^2 ) +  n^2 ( 10 + R )  }
  {2\,( 6 + R )  - n\,( 14 + 13 R + R^2 )  + n^2 ( 22 + 13 R + R^2 ) }.
\eeq

In the following, we consider models corresponding   to Markov processes with rates which are
piece-wise constant functions
of time $t$. This allows us to calculate
$\langle\tau_{x(ij)}\tau_{y(ij)}\rangle$ from (\ref{eq:corr}) by taking
$\mathbf{M}$ and ${\bm u}$ to be functions of time.

\clearpage \newpage %
\section{Results}
\label{sec:results}

After having illustrated our approach, we now briefly describe
the demographic models we have considered and summarise our results
for gene-history correlations in these models. Mathematical details
are given in appendices A and B. Implications are discussed
in section 5.

\subsection{Bottleneck model}
\label{sec:bottleneck_model}

Consider (c.f.~\cite{eyre-walker_etal98}) an unstructured
population of constant size $N$ until $\tau_0 = 2 N G$ generations
ago. The population was then subject to a severe bottleneck of
short duration, followed by a rapid expansion to a very large
(infinite) population size (figure~\ref{fig:pop struct models}a).
Between the bottleneck and now, the population size is taken to be
effectively infinite: and thus the probability that two randomly
sampled individuals have a common ancestor before the bottleneck
is negligible. Since the bottleneck is very narrow and has a short
duration, we may ignore the effect of recombination during the
bottleneck. It is convenient to parameterise the duration of the
bottleneck in terms of the probability $F$ that a single locus
coalesces during the bottleneck. In the limit when both the
population size and duration of the bottleneck are small (compared
to $2N$ individuals and generations, respectively), we obtain
(appendix A):
\begin{equation}\label{eq:rho_bottleneck}
   \rho(\tau_{x(ij)},\tau_{y(ij)})  = \frac{A + B\,e^{-R G/2} + C\,e^{-R G}}{15\,
   (2 - h)\,(18 + 13\,R + R^2 ) }\,,
\end{equation}
where $h = 1 - F$ and
\begin{eqnarray}
   A &=& 6 ( 36 - 45 h + 20 h^2 - h^5 )
                + 3 ( 28 - 65 h +\nonumber\\&&+\ 40 h^2 - 3 h^5 ) R
                + {( 1 - h ) }^3 ( 6 + 3 h + h^2) R^2 \,, \\
   B &=& 12( 9 - 5 h^2 + h^5 ) + ( 3 - 5 h^2 + 2 h^5 ) R^2\nonumber\\
               &&+\ 6 ( 7 - 10 h^2 + 3 h^5 ) R \,, \\
   C &=& 6 ( 36 - 10 h^2 - h^5 ) + ( 6 - 5 h^2 - h^5 ) R^2  \nonumber\\
               &&+\ 3 ( 28 - 20 h^2 - 3 h^5 ) R \,.
\end{eqnarray}
We thus find
that this model exhibits correlations at arbitrarily large values of $R$, 
a consequence of an infinite expansion rate after the bottleneck,
and negligible recombination within it. If, instead, the expansion
were to a finite population size, (smaller than $GN$, say), the
correlations would still converge to a constant at large $R$. The
constant, however, is expected to be lower than the asymptotic
value obtained from (4) as $R\rightarrow\infty$. Finally, if the
bottleneck lasts long enough for significant recombination to
occur within it, we still find long-range correlations, up to
scales of the order of $(2\tau_{\rm D}r)^{-1}$ where $\tau_{\rm
D}$ is the duration of the bottleneck (in generations). Beyond
this, the correlations decay, and in the limit $R\rightarrow\infty$
we have $\rho(\tau_{x(ij)},\tau_{y(ij)})\rightarrow 0$ as in the
unstructured population model.

By the same approach, we calculate
$\expt{\tau_{x(ij)}\tau_{y(ik)}}$ and $\expt{ \tau_{x(ij)}
\tau_{y(kl)}}$. Inserting this into (\ref{eq:sigma2 def}) yields,
for large $n$:
\begin{eqnarray}\label{eq:sigma2_bottleneck}
   \sigma^2_d &=& \frac{e^{-G\,R}}{\expt{ \tau_{x(ij)} \tau_{y(kl)}}} \Big[ 18\,h\,( 36 - 10\,h^2 - h^5)  +
   9\,h\,( 28 - 20\,h^2 - 3\,h^5) \,R + \nonumber\\&& 3\,h\,( 6 - 5\,h^2 - h^5) \,R^2 \Big] \, ,
\end{eqnarray}
where
\begin{eqnarray}
  \expt{ \tau_{x(ij)} \tau_{y(kl)}} &=& 18\,( 45\,G^2 + 36\,h + 90\,G\,h + 20\,h^3 - h^6 )  +\nonumber\\&&
  9\,( 65\,G^2 + 28\,h + 130\,G\,h + 40\,h^3 - 3\,h^6) \,R +\nonumber\\&&
  ( 45\,G^2 + 18\,h + 90\,G\,h + 30\,h^3 - 3\,h^6) \,R^2 \, .
\end{eqnarray}
Note that $\sigma^2_d \rightarrow 0$ as $R \rightarrow \infty$.
The difference, in particular, to expression (7) is not large.
Hence, when the aim is to detect the population-size variations it
is better to focus on single-locus statistics.

\subsection{Model of divergent populations, I}
\label{sec:div_model_1}

Reich {\em et al.} consider a model of a diverging population
\cite{reich_etal02}: the population was unstructured with constant
population size $N$ until $\tau_0 =2 N G$ generations ago, when
the the population split into two parts of equal size $N$ (note
that this implies a rapid population expansion from $N/2$ to $N$
after the split). The model is illustrated in figure~\ref{fig:pop
struct models}c. A portion $p$ of the sample is chosen from the
first population, and the rest from the second population. For any
two individuals in the sample, the expectation
$\rho(\tau_{x(ij)},\tau_{y(ij)})$ depends on whether the
individuals come from the same sub-population or not. Using the
technique illustrated above, it is straightforward to calculate
the expectation for both cases. Again, we find long-range
correlations, namely
\begin{equation}\label{eq:corr_model_2c}
   \rho(\tau_{x(ij)},\tau_{y(ij)})
   = 1 - \frac{1}{1 + 2\,p\,(1-p)\,(1 - 2\,p + 2\,p^2)\,G^2} \,,
\end{equation}
in the limit of large $R$ (in appendix B we describe how to
obtain the full result, valid for arbitrary values of $R$).

Further, in the limit of large $R$ and large sample size $n$, we have
\beq\label{eq:sigma2_model_2c}
   \sigma^2_d = \frac{2\, p^2\,(1 - p)^2\,G}{ 1 + 2\,p\,(1-p)\,G} .
\eeq
Thus, for this model $\sigma^2_d$ is finite in the limit of large
$R$, as opposed to $\sigma^2_d$ in the unstructured model (section
\ref{sec:gene-history correlations}) and the bottleneck model
(section \ref{sec:bottleneck_model}).

\subsection{Model of divergent populations, II}
\label{sec:div_model_2}

Now consider the model of two diverging sub-populations
\cite{eyre-walker_etal98} in figure~\ref{fig:pop struct models}b.
The population was unstructured with constant size of $N$
individuals until $\tau_0=2 N G$ generations ago, when a fraction
$\gamma$ of the population diverged. In subsequent generations,
the two sub-populations where unstructured but with no contact
between sub-populations. Individuals are randomly chosen from the
joint population. For two individuals in the sample, there are
three cases: both individuals may come from the smaller
sub-population, they may come from the larger sub-population, or
from different sub-populations. Using equation (\ref{eq:corr}) we
find long-range correlations: in the limit of large $R$,
$\rho$ remains finite,
\begin{eqnarray}\label{eq:corr_model_2b}
   \rho(\tau_{x(ij)},\tau_{y(ij)})  &=& \frac{1}{\mbox{var}[\tau]} \big[
   1 - 2 s + 2 s^2 + 2 G\left( 2 + G \right) s +
   s^2  {\rm e}^{-\frac{2 G}{\gamma }} +\\
   &&s^2 {\rm e}^{-\frac{2 G}{1 - \gamma }} +
   2 s {\left( 1 - \gamma  \right) }^2  {\rm e}^{-\frac{G}{1 - \gamma }} +
   2 s {\gamma }^2  e^{-\frac{G}{\gamma }}  - \left<\tau\right>^2
   \big]\,
   \nonumber
\end{eqnarray}
where $s = \gamma\,(1-\gamma)$ and
\begin{eqnarray}
   \left<\tau\right> &=& 1 + s (2 G - 1) + s \gamma {\rm e}^{-\frac{G}{\gamma}}
   + s (1 - \gamma) {\rm e}^{-\frac{G}{1 - \gamma}}\, \\
   \mbox{var}[\tau] &=&  2 + 2 s  \big[ 2 s  + (G + 1)^2 +
       \gamma (1 + G + \gamma) {\rm e}^{-\frac{G}{\gamma}}
       +\nonumber\\&&+\ (1 - \gamma) (2 + G - \gamma)
       {\rm e}^{-\frac{G}{1 - \gamma}} - 3 \big] - \left<\tau\right>^2\,.
\end{eqnarray}
See the appendix for the full result. The long-range correlations
are found to be due to sampling of different sub-populations.

In the limit of large $R$ and large sample size, we have
\beq\label{eq:sigma2_model_2b}
   \sigma^2_d = \frac{\gamma^2 (1 - \gamma)^2}{\expt{\tau}^2} \left[ 2\,G + \gamma\,(1 - \rme^{-\frac{G}{1-\gamma}}) + (1 - \gamma)(1 - \rme^{-\frac{G}{\gamma}}) \right]^2 .
\eeq
Again, we find that $\sigma^2_d$ is finite in the limit of large
$R$.

\clearpage \newpage %
\section{Discussion}
\label{sec:discussion}

\begin{figure}
   \centerline{\includegraphics{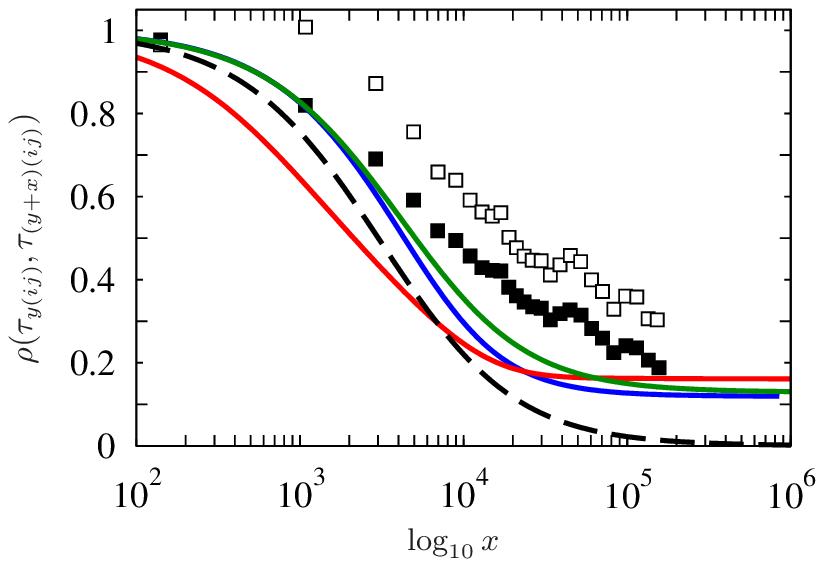}}
\caption{\label{fig:pop struct results}
Correlation $\rho(\tau_{y(ij)},\tau_{(y+x)(ij)})$ of gene histories as a
function of the distance $x$ between them. Equations
(\ref{eq:rho_no_pop_struct}), (\ref{eq:rho_bottleneck}), and exact
expressions corresponding to (\ref{eq:corr_model_2c}) and
(\ref{eq:corr_model_2b}), from the appendix, were used. In all
cases, $r = 1.2$ cM/Mb, $N$ and $\mu$ were chosen to be
consistent with $2N\left<\tau\right> = 1.55\times 10^4$, and a
coefficient of variation of $0.94$ \cite{reich_etal02} (except in
the unstructured model). The lines are: the unstructured
coalescent (dashed), bottleneck model with $H = 0.1$ (red),
divergent model in figure~\ref{fig:pop struct models}b with
$\gamma = 0.2$ (blue), and divergent model in figure~\ref{fig:pop
struct models}c with $p = 0.3$ (green). Also shown are empirical
estimates of lower and upper bounds for the correlation of gene
histories in the human genome (squares) \cite{reich_etal02}.
}
\end{figure}

Figure~\ref{fig:pop struct results} shows the correlations
$\rho(\tau_{x(ij)},\tau_{y(ij)})$ in the demographic models
considered, with parameters chosen to be consistent with the
empirically estimated time to the most recent common ancestor and
its coefficient of variation \cite{reich_etal02}.
When plotting the correlation of gene histories against physical
positions, we need to translate the recombination fraction $r$
into the corresponding expected number $\sigma x$ of crossover
events between the two loci. There are many such maps proposed in
the literature (see e.g. \cite{mcpeek_speed95} for a review of
these). They differ in how they model the chiasma process, but all
models have in common that for small enough $r$, $r \approx \sigma
x$. In humans, $r \approx \sigma x$ for $x \lesssim 10^6$bp. At
larger distances, deviations from linearity are not noticeable
since the expressions for $\rho(\tau_{x(ij)},\tau_{y(ij)})$ and
$\sigma^2_d$ converge for large $R$ (to different values, in
general).
Also shown are empirical estimates of lower and upper bounds on
the correlation of gene histories in the human genome
\cite{reich_etal02}. The correlations for the models described in
section \ref{sec:results} are substantially larger at large
distances than those for the unstructured model, but they lie
significantly below the lower bound of the empirical data, at
intermediate distances. We comment on possible causes for
this discrepancy in our conclusions.

Our results allow us to gain a qualitative understanding
of the influence of demographic factors on the decorrelation
of gene histories.
First, we find that models of bottlenecks and divergent
populations  (figure~\ref{fig:pop struct models}) both exhibit
long-range correlations in gene histories, as numerically
demonstrated in \cite{reich_etal02}, but for very different
reasons. In bottlenecks, the length scale at which we find
significant correlations is governed by the degree of
recombination
within the
bottleneck: low recombination in the bottleneck gives rise to
long-range correlations. Further, the amount of correlation is
affected by the rate of expansion of the population after the
bottleneck: rapid expansion gives high correlations. Long-range
correlation in divergent models, on other hand, we ascribe to the
fact that the covariance of $\tau_{x(ij)}$ and $\tau_{y(ij)}$ (that is, the
number of generations since the common ancestor of two copies of
loci $x$ and $y$) is different when individuals are selected from
the same or different sub-populations: typically, the covariance
is lower for individuals from the same sub-population than from
different ones. We find that this effect persists even for loci
far apart, but is decreased by population expansions during the
divergence.

Second, we identify two contributions to the correlation of gene
histories in divergent populations: linkage disequilibrium and the
sampling of sub-populations with different demographic histories.
At short ranges, linkage disequilibrium correlates nearby patterns
by co-inheritance. Thus, for small distances, we conclude that the
demographic structure is unimportant: all reasonable models must
give high correlation for small distances. For long ranges, by
contrast, correlations due to linkage disequilibrium are expected
to vanish, but the contribution from differences in gene history
across sub-populations remains.

Third, the domestication of crops and animals has shaped the
genetic makeup of the species, through selection for desirable
traits but also through the demographic history of each species
\cite{eyre-walker_etal98}. The pattern of genetic differences in
the laboratory mouse population depends strongly on its
demographic history \cite{wade_etal02}. In divergent populations,
we find that long-range correlations are insensitive to the
demographic history of the sub-populations. As a consequence, we
predict that the most important contribution to the correlation of
gene history in the laboratory mouse is from the original
divergence from the wild-type mouse.

Fourth, we found that within the models described in
section~\ref{sec:results}, gene-history correlations are
substantially increased as compared with the unstructured,
standard model. However, the correlations still lie significantly
below the empirically determined data at intermediate distances.
In \cite{eriksson_mehlig04} it was shown that incorporating
empirically observed variations in the recombination-rate along
the chromosomes \cite{kong_etal02} significantly increases the
correlations in this regime.
Our analytical expressions for the correlation of gene
histories allow for studying the effect  of such variations in the
recombination rate in models with demographic population structure.

Fifth,  we briefly mention possible extensions of the scheme introduced
in this paper.
In more general sampling schemes (different from those depicted
in figure~\ref{fig:pop struct models}), we may use the expressions for
$\left<\tau_{x(ij)}\,\tau_{y(ij)}\right>$ conditional on whether the
individuals in the sample came from the same sub-population or
not, and conditional on the population size during the divergence,
to calculate the correlation of gene histories by weighting the
different contributions by the probability that they occur under
the sampling scheme. Also, it is straight-forward to extend the
calculations to combinations of bottlenecks and divergent
populations (figure~\ref{fig:pop struct models}d), and to more
complicated models involving more than two diverging branches
(figure~\ref{fig:pop struct models}e). It is expected that the
most distant (symmetric) divergence determines the long-range
correlations.

How would a recent mixing event (figure~\ref{fig:pop struct
models}e) affect the correlation of gene histories? A merging of
the divergent populations $g$ generations ago leads to a
decorrelation of gene histories at distances of the order of $(4 g
r)^{-1}$, since then ancestral lines of both loci may come from
different sub-populations with approximately equal probability.

\begin{figure}
   \centerline{\includegraphics{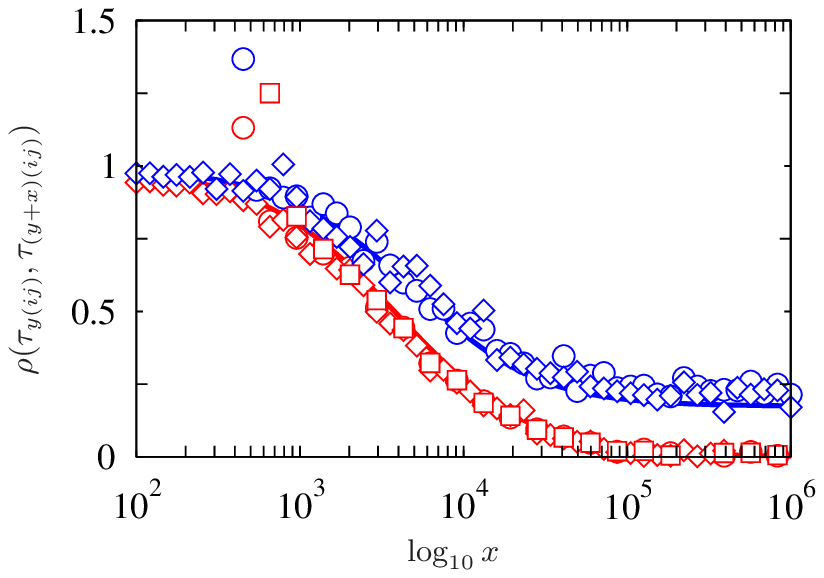}}
\caption{\label{fig:estimate_rho} %
Comparison of $\hat{\rho}(x)$ (markers) to
$\rho(\tau_{y(ij)},\tau_{(y+x)(ij)})$ (solid lines, calculated from theory),
for an unstructured population (red) and a divergent population
(blue). The estimator $\hat{\rho}(x)$ were obtained from a single
sample of 50 individuals, with $L_\mathrm{c} = 10$Mb, for
different bin sizes $L = 100$bp (diamonds), $L = 500$bp (circles)
and $L = 1$kb (squares). The parameters for the divergent model
are: $G = 0.6$, $p = 0.3$, $N = 6963.7$, $r = 0.95633$cM/Mb,
$\theta = 7.6\,10^{-4}$. In the unstructured population model, the
population size is $N = 10^4$.
}
\end{figure}

Finally, we have argued that the correlation $\rho(\tau_{x(ij)},\tau_{y(ij)})$ of gene
histories determines the association of SNP counts,
$\mbox{cov}[S_{x(ij)},S_{y(ij)}]$. Conversely one may be interested
in estimating model
parameters from population data, deducing
$\rho(\tau_{x(ij)},\tau_{y(ij)})$
from the pairwise statistic $\mbox{cov}[S_{x(ij)},S_{y(ij)}]$.
Three questions arise. First, how can one in practice estimate $\mbox{cov}[\tau_{x(ij)},\tau_{y(ij)}]$
from              the variance of SNP counts? Second,
how good is this estimate? Third, how much of
the information the full data set (possibly pertaining to a large
number of individuals) is retained in the pair-wise statistic
$\mbox{cov}[S_{x(ij)},S_{y(ij)}]$?
We begin by answering the last question.
Due to the high amount of association between the chromosomes in a
sample, the information on genealogical history accumulates slowly as the
sample size is increased \cite{hudson01}. It follows that most
information can be found in pair-wise comparisons between the
chromosomes in the sample as used in eq.~(\ref{eq:cov S_a S_b}).
Going back to the first two questions, an estimator for
$\rho(\tau_{y(ij)},\tau_{(y+x)(ij)})$ can be
constructed as follows.
Assuming that the length $L_\mathrm{c}$
of the sequences is long, we can estimate the correlation of
polymorphism rates by averaging over all pairs and positions:
\begin{equation}\label{eq:estimate_rho}
   \rho(\tau_{y(ij)},\tau_{(y+x)(ij)}) \approx \hat{\rho}(x) =  \frac{\overline{S_y S_{y+x}} - \overline{S_y}^2}{\overline{S_y^2} - \overline{S_y}^2  - \overline{S_y}},
\end{equation}
where
\begin{equation}\label{eq:sequence_average_def}
   \overline{S_y S_{y+x}} = \frac{2}{n(n-1)(L_\mathrm{c} - x - L)}
   \sum_{i=2}^n \sum_{j=1}^{i-1} \sum_{y=1}^{L_\mathrm{c}-x-L} S_{y(ij)} S_{(y+x)(ij)} \,.
\end{equation}
and the single-locus quantities $\overline{S_y}$ and
$\overline{S_y^2}$ are defined similarly. Instead of regularly
spaced bins, as in (\ref{eq:sequence_average_def}), one may use
randomly positioned bins. For unstructured populations, and for
populations with bottlenecks and expansions, the accuracy of the
estimator $\hat{\rho}(x)$ depends mostly on the number of bins
(and hence on $L_\mathrm{c}$), and improves only slowly with
increasing $n$. For divergent models, however, increasing $n$
improves the sampling from the different sub-populations. In
figure~\ref{fig:estimate_rho} we show how $\hat{\rho}(x)$ compares
to $\rho(\tau_{y(ij)},\tau_{(y+x)(ij)})$ when applied to a sample. As can be
seen in the figure, when $x < L$ the bins overlap and
$\hat{\rho}(x)$ overestimates the correlations, but
otherwise it works well.

\clearpage \newpage %
\section{Conclusions and outlook}
\label{sec:conclusions}

We have derived closed analytical expressions for the correlation
of gene histories in established demographic models for genetic
evolution. These expressions allow us to understand and quantitatively
determine how demographical factors give rise to long-range
correlations in gene histories.

The correlations analysed here determine
the two-person summary statistic (\ref{eq:cov S_a S_b}). 
More information is contained in the mosaics of SNP
haplotype patterns for more than two individuals, and their
associations \cite{hudson01}. It is of great interest to derive
corresponding expressions for correlations between such patterns
in the models considered in this paper, especially
in the case of more than two loci.
Finally we note that the
quantity $\sigma_d^2$, a measure of linkage disequilibrium, was
shown to be a good approximation to $r^2$ in the case of
unstructured populations \cite{mcvean_etal02}. It is necessary to
investigate the relation between $r^2$ and $\sigma^2_d$ in models
with demographic structure.

\clearpage \newpage %
{ \appendix

\section*{Appendix A: Derivation of bottleneck formula}
\setcounter{section}{1}
\label{app:bottleneck formula}

During the bottleneck, the time between coalescent events is
exponentially distributed with rate ${n \choose 2}/(2\,\gamma N)$,
where $n$ is the number of lines carrying ancestral material.
Recombination events occurs with rate $n\,R/(4N)$, independent of
$\gamma$. Thus when $\gamma$ is very small, coalescent events
dominate the process.

We assume that during the bottleneck, the reduction in effective
population size is so drastic that $\gamma$ is effectively zero.
By rescaling the time by a factor of $\gamma$ and taking the limit
of $\gamma \rightarrow 0$ we find
\be
   \mathbf{M}' = \lim_{\gamma \rightarrow 0} \mathbf{M}(\gamma)\,\gamma =
   \left[\begin{array}{rrr}
         -1 & 1 & 0 \\
         0 & -3 & 4 \\
         0 & 0 & -6
   \end{array}\right],
\ee
so the time evolution operator becomes
\be
   \exp(\mathbf{M}'\,t) =
   \left[\begin{array}{rrr}
      e^{-t} &  \frac{1}{2}\,e^{-t} - \frac{1}{2}\,e^{-3 t} & \frac{2}{5}\,e^{-t} - \frac{2}{3}\,e^{-3 t} +  \frac{4}{15}\,e^{-6 t}\\
      0 & e^{-3 t} &  \frac{4}{3}\,e^{-3 t} - \frac{4}{3}\,e^{-6 t}\\
      0 & 0 & e^{-6 t} \\
  \end{array}\right] .
\ee
In the original model, the inbreeding coefficient $F$ was
specified. We choose to parameterise the severity of the
bottleneck by its duration $D$. If the process is in state $1$
(figure~3)
when entering the bottleneck, the probability of coalescence
during the bottleneck is
\be
   \int_0^D {\bm u}_1\transpose\,\rme^{\mathbf{M}'\,t}\,{\bm u}_1 \,\rmd t = 1 - \rme^{-D},
\ee
so we see that by taking $D = -\ln(1 - F)$, we get the correct
inbreeding coefficient. We can now express the time evolution
operator from the beginning to the end of the bottleneck as
\be\label{eq:propagator_bn}
   \exp(\mathbf{M}'\,D) =
   \left[\begin{array}{rrr}
      H & \frac{1}{2}\,H\,(1 - H^2) & \frac{2}{15}\,H\,( 3 - 5\,H^2 + 2\,H^5 ) \\
      0 & H^3 & \frac{4}{3}\,H^3\,(1 - H^3) \\
      0 & 0 & H^6
   \end{array}\right] ,
\ee
where $H = 1 - F$. The probability that the loci become linked
during the bottleneck depends on the state of the process when the
bottleneck is entered:
\be\label{eq:prob_linked_bn}
   \int_0^D {\bm u}_1\transpose\, \rme^{\mathbf{M}'\,t} \,\rmd t =
   \left\{\begin{array}{ll}
      F & \mbox{in state $1$} \\
      \frac{1}{6}\,( 2 + H )\,F^2 & \mbox{in state $2$} \\
      \frac{2}{45}\,( 5 + 6\,H + 3\,H^2 + H^3 )\,F^3 &  \mbox{in state $3$}
   \end{array}\right.
\ee
Similarly, we have the probability that one locus, but not the
other, reaches its most recent common ancestor during the
bottleneck, depending on the state of the process when entering
the bottleneck:
\be\label{eq:prob_apart_bn}
   \int_0^D {\bm u}_2\transpose\, \rme^{\mathbf{M}'\,t} \,\rmd t =
   \left\{\begin{array}{ll}
      0  & \mbox{in state $1$}\\
      \frac{2}{3} \,( 1 - H^3) & \mbox{in state $2$} \\
       \frac{1}{9}\,(7 - 8\,H^3 + H^6) & \mbox{in state $3$}
   \end{array}\right.
\ee
Together, (\ref{eq:propagator_bn}), (\ref{eq:prob_linked_bn}) and
(\ref{eq:prob_apart_bn}) determines the
state of the process after the bottleneck. Using this
information and the method
for the unstructured population as outlined in section 2 allows
us to derive the gene-history correlation for the bottleneck model.

\section*{Appendix B: Correlation of gene histories in divergent
populations}
\setcounter{section}{2}
\label{app:div pop}

Assume that individuals come from left sub-population with
probability $p$ and from the right one with probability $1-p$. The
population size in the left and right sub-populations are $\gamma
N$ and $\Gamma N$, respectively, and the population size before
the divergence is $N$.
The two-person coalescent process is described by a Markov process
over the states in table~\ref{tab:states}, where state $1$ is the
absorbing state of the process, and the process starts in one of
states $3 - 11$.

\begin{table}
\caption{\label{tab:states} %
The states of the Markov process of loci $x$ and $y$ in
chromosomes $i$ and $j$, for the divergent population. For each
state we show the corresponding configurations of the
sub-populations, separated by a vertical bar. A dash denotes
genetic material that is not ancestral to any locus in the sample.
The symbol $\phi$ denotes a sub-population unrelated to sample,
and the diamonds denotes a common ancestor to chromosomes $i$ and
$j$ (for that locus).
}
\begin{indented}
   \item[]\begin{tabular}{c r@{ $|$ }l}
   \br
   State  & \multicolumn{2}{c}{Population configuration} \\
   \mr
   0 & $\phi$ & $\phi$ \\
   \mr
   1 & $x_i \diamond$, $x_j \diamond$ & $\phi$ \\
   2 & $x_i \diamond$ &  $x_j \diamond$ \\
   \mr
   3 & $x_i y_i$, $x_j y_j$ & $\phi$ \\
   4 & $x_i y_i$ & $x_j y_j$ \\
   \mr
   5 & $x_i y_i$, $x_j-$, $-y_j$ & $\phi$ \\
   6 & $x_i y_i$, $x_j-$ & $-y_j$ \\
   7 & $x_i y_i$ & $x_j-$, $-y_j$ \\
   \mr
   8 & $x_i-$, $-y_i$, $x_j-$, $-y_j$ & $\phi$ \\
   9 & $x_i-$, $-y_i$, $x_j-$ & $-y_j$ \\
   10 & $x_i-$, $-y_i$ &  $x_j-$, $-y_j$  \\
   11 & $x_i-$, $x_j-$ & $-y_i$, $-y_j$ \\
   \br
\end{tabular}
\end{indented}
\end{table}
We now define $e_i = \expt{\, \ta \tb\, |\, \mbox{Process starting
in state $i$}\,}$. With these, we may write
\be
   \expt{ \tau_{x(ij)} \tau_{y(ij)} } &=&
         p^2\, e_3(\gamma) + (1 - p)^2\, e_3(\Gamma) + 2 p (1 - p)\, e_4(\gamma,\Gamma), \\
   \expt{ \tau_{x(ij)} \tau_{y(ik)} } &=&
         p^3\, e_5(\gamma) + (1-p)^3\, e_5(\Gamma) \nonumber\\&+&
         2 p (1 - p)^2\, e_6(\gamma)  + 2 p^2 (1 - p)\, e_6(\Gamma) \nonumber\\&+&
         p (1 - p)^2\, e_7(\gamma, \Gamma) + p^2 (1 - p)\, e_7(\Gamma, \gamma), \\
   \expt{ \tau_{x(ij)} \tau_{y(kl)} } &=&
         p^4\, e_8(\gamma) + (1 - p)^4\, e_8(\Gamma) \nonumber\\&+&
         4 p^3 (1 - p)\, e_9(\gamma) + 4 p (1 - p)^3\, e_9(\Gamma) \nonumber\\&+&
         4 p^2 (1 - p)^2\, e_{10}(\gamma,\Gamma) + 2 p^2 (1 - p)^2\, e_{11}(\gamma,\Gamma).
\ee
From this, the correlation $\rho(\tau_{x(ij)},\tau_{y(ij)})$ and $\sigma^2_d$
may be calculated for both models of divergent populations:
setting $\gamma = \Gamma = 1$ gives the model described in section
\ref{sec:div_model_1}; setting $\Gamma = 1 - \gamma$ and $p =
\gamma$ gives the model described in section
\ref{sec:div_model_2}.

\subsection*{Calculation of $e_3,\ldots,e_{11}$ for the model
introduced in section 4.2}

\newcommand{\MM}{\mathbf{M}_1}

The two-locus coalescent in a population of size $\gamma N$ is
described by a Markov process with the evolution matrix
\beq
  \MM = \left[ \begin{array}{ccc}
       - 1/\gamma - R & 1/\gamma & 0  \\
      R &  - 3/\gamma - R/2 & 4/\gamma  \\
      0 & R/2 &  - 6/\gamma
   \end{array}  \right]\!\!.
\eeq
where $R = 4Nr$. Before the divergence, $\gamma = 1$ and we denote
the corresponding evolution matrix $\mathbf{M}$. the coalescent is
described by a Markov process with the evolution matrix
$\mathbf{M}$. Assuming that population is in state $3$, $5$, or
$8$ with probabilities $v_1$, $v_2$, and $v_3$, respectively, we
proceed as for the unstructured population in section
\ref{sec:methods}, calculating $\expt{\ta \tb}$ conditional on
starting from distribution ${\bm v}$. We obtain
 $e_3(\gamma) = c_\mathrm{s}(\gamma, (1,\, 0,\, 0)\transpose)$,
 $e_5(\gamma) = c_\mathrm{s}(\gamma, (0,\, 1,\, 0)\transpose)$,
 and
 $e_8(\gamma) = c_\mathrm{s}(\gamma, (0,\, 0,\, 1)\transpose)$,
where
\be
 \fl    c_\mathrm{s}(\gamma, {\bm v}) &=& \frac{{\bm u}_1\transpose}{\gamma} \, (-\MM)^{-3} \, \big[2\,\mathbf{I} - (2\,\mathbf{I} - 2\,\frac{G}{\gamma}\,\MM + \frac{G^2}{\gamma^2}\, \MM^2)\,\expm{\MM G}\big] {\bm v} \nonumber\\
 \fl    &+& {\bm u}_1\transpose \, (-\mathbf{M})^{-3} \, \big(2\,\mathbf{I} - 2\,G\,\mathbf{M} + G^2\,\mathbf{M}^2 \big) \, \expm{\MM G}  {\bm v} \nonumber\\
 \fl    &+& \frac{ {\bm u}_2\transpose}{\gamma} \, (-\MM)^{-3}  \Big\{ 2\,\mathbf{I} - \gamma\,\MM - \left[ 2\,\mathbf{I} - (2\,G + \gamma)\,\MM + G\,(G + \gamma)\, \MM^2 \right]\expm{\MM G} \Big\}  {\bm v} \nonumber\\
 \fl    &+& (1 - \gamma)\,{\bm u}_2\transpose \, (\mathbf{I} + \gamma\,\MM)^{-2} \, \Big\{ \gamma\,e^{-G/\gamma}\,\mathbf{I} + \left[ \, (G - \gamma)\,\mathbf{I} + \gamma\,G\,\MM \, \right]\,\expm{\MM G} \Big\}  {\bm v} \nonumber\\
 \fl    &+& {\bm u}_2\transpose (-\mathbf{M})^{-3} \left[ 2\,\mathbf{I} - (1 + 2\,G)\,\mathbf{M} + G\,(G + 1)\,\mathbf{M}^2 \right] \expm{\MM G}  {\bm v} .
\ee

During the split, the coalescent is described by a Markov process
with the evolution matrix
\beq
  \mathbf{M}_2 = \left[ \begin{array}{cc}
      - 1/\gamma - R/2 & 2/\gamma  \\[2pt]
      R/2 &  - 3/\gamma
   \end{array}  \right]\!\!.
\eeq
A coalescent event during the split happens with the distribution
 $\gamma^{-1} (1,\, 1)\, \rme^{\mathbf{M}_2 \ta} {\bm v},$
where ${\bm v} = (1,\, 0)$ when starting from state $6$ and ${\bm
v} = (0,\, 1)$  when starting from state $9$. Thus, we have the
contribution
\[
 \int_0^G\!\! \ta \, \frac{1}{\gamma}\, (1,\, 1) \rme^{\mathbf{M}_2 \ta}\, {\bm v}\, \rmd\ta
 \int_G^\infty\!\! \tb\, \rme^{-(\tb - G)} \rmd\tb
\]
The population is in state $5$ or $8$, right before the split,
with probability ${\bm a}\, \expm{\mathbf{M}_2 G}\, {\bm v}$,
where ${\bm a} = (1, 0)$ for state $5$ and ${\bm a} = (0, 1)$ for
state $8$. From this we obtain
\be
   e_6(\gamma) &=& A(\gamma) + R \gamma \,  B(\gamma) \nonumber\\
   e_9(\gamma) &=& A(\gamma) - 2\, B(\gamma) \nonumber
\ee
where
\beq
 \fl  A(\gamma) = (1 + G) \gamma + \left[ (1 + G)(1 - \gamma)  + \frac{24 + 4 R \gamma }{( 4 + R \gamma)( 18 + 13 R + R^2 ) } \right] \mathrm{e}^{-G/\gamma}
\eeq
and
\beq
 \fl  B(\gamma) = \frac{2}{( 4 + R\,\gamma) \, ( 18 + 13\,R + R^2 ) }\, \exp\!\left(- \frac{G\,( 6 + R\,\gamma) }{2\,\gamma }\right)
\eeq

Now consider starting from states $4$, $7$ or $10$.
In these cases, there is no coalescent event during the split. In
each sub-population the coalescent is described by a Markov
process with the evolution matrix
\beq
  \mathbf{M}_3 = \left[ \begin{array}{cc}
      - R/2 & 1/\gamma  \\[2pt]
      R/2 &  - 1/\gamma
   \end{array}  \right]\!\!.
\eeq
Note that the columns sum to zero: the probability of escaping
from these states is zero during the split.

Right before the split, the population is in state $3$, $5$ or $8$
with probability $\phi_1$, $\phi_2$, and $\phi_3$, respectively.
Then, the contribution is
\be
 \fl  && \int_G^\infty \! \left[ \ta^2\, {\bm u}_1\transpose + \int_{\ta}^\infty\!\! \ta \tb\, \rme^{\ta - \tb}\, \rmd\tb \, {\bm u}_2\transpose \right] \rme^{\mathbf{M}\,(\ta - G)}\, {\bm \phi} \,\, \rmd\ta \nonumber\\
 \fl  &&\hspace{1cm}=\ (1 + G)^2 (\phi_1 + \phi_2 + \phi_3)  + \frac{(R + 18)\phi_1 + 6 \phi_2 + 4 \phi_3}{R^2 + 13 R + 18}
\ee
Now define $P_\mathrm{L}(\gamma)$ as the probability of the
genetic material being on the same gamete at the moment of the
split, given that it is on the same gamete in the sample. We have
\beq
   P_\mathrm{L}(\gamma) = (1,\, 0)\, \expm{\mathbf{M}_3\, G}\, (1,\,0)\transpose = \frac{2 +  R \gamma \exp\!\left(- \frac{G (2 + R \gamma)}{2 \gamma} \right)}{2 + R \gamma}.
\eeq
Similarly, we define $P_\mathrm{B}(\gamma)$ as the probability of
the genetic material being on the same gamete at the moment of the
split, given that it is on different gametes in the sample. We
have
\beq
   P_\mathrm{B}(\gamma) = (1,\, 0)\, \expm{\mathbf{M}_3\, G}\, (0,\,1)\transpose = \frac{2 -  2  \exp\!\left(- \frac{G (2 + R \gamma)}{2 \gamma} \right)}{2 + R \gamma}.
\eeq
If the sample is in state $4$, we have
\be
   \phi_1 &=& P_\mathrm{L}(\gamma) \, P_\mathrm{L}(\Gamma) \nonumber\\
   \phi_2 &=& P_\mathrm{L}(\gamma) \, [1 -  P_\mathrm{L}(\Gamma)] + [1 - P_\mathrm{L}(\gamma)] \, P_\mathrm{L}(\Gamma) \nonumber\\
   \phi_3 &=& [1 - P_\mathrm{L}(\gamma)] \, [1 - P_\mathrm{L}(\Gamma)]
\ee
Since $\phi_1 + \phi_2 + \phi_3 = 1$ we have
\beq
  \fl e_4(\gamma,\Gamma) = (1 + G)^2 + \frac{4 + 2\,P_\mathrm{L}(\gamma) + 2\,P_\mathrm{L}(\Gamma) + (10 +  R)\,P_\mathrm{L}(\gamma)\, P_\mathrm{L}(\Gamma) }{R^2 + 13 R + 18}
\eeq
Similarly, we obtain
\beq
  \fl  e_7(\gamma,\Gamma) = (1 + G)^2 + \frac{4 + 2\,P_\mathrm{L}(\gamma) + 2\,P_\mathrm{B}(\Gamma) + (10 +  R)\,P_\mathrm{L}(\gamma)\, P_\mathrm{B}(\Gamma) }{R^2 + 13 R + 18}
\eeq
and
\beq
  \fl  e_{10}(\gamma,\Gamma) = (1 + G)^2 + \frac{4 + 2\,P_\mathrm{B}(\gamma) + 2\,P_\mathrm{B}(\Gamma) + (10 +  R)\,P_\mathrm{B}(\gamma)\, P_\mathrm{B}(\Gamma) }{R^2 + 13 R + 18}
\eeq
%
%
Finally, starting from state $11$, we obtain
\beq
  \fl e_{11}(\gamma,\Gamma) =
   \frac{4}{18 + 13R + R^2} \, \rme^{-G/\gamma - G/\Gamma} +
   \left[ \gamma  + ( 1 - \gamma  ) \mathrm{e}^{-G/\gamma} \right]\!
   \left[ \Gamma  + ( 1 - \Gamma  )\mathrm{e}^{-G/\gamma} \right]
\eeq

\subsection*{Calculation of $e_3,\ldots,e_{11}$ for the 
model introduced in section 4.3}
In this model, $\gamma = \Gamma = 1$ so the formulas simplify
considerably. Starting from state $3$, $5$ or $8$, we obtain
\be
   e_3 &=& 1 + \frac{18 + R}{R^2 + 13 R + 18} \nonumber\\
   e_5 &=& 1 + \frac{6}{R^2 + 13 R + 18} \nonumber\\
   e_8 &=& 1 + \frac{4}{R^2 + 13 R + 18} \nonumber\\
\ee
as calculated by Griffiths \cite{griffiths81}. Starting from state
$6$ or $9$, we obtain
\be
   e_6 &=& (1 + G)^2 + \frac{ (24 + 4 R) \rme^{-G} + 2 R\, \rme^{-G(6 + R)/2}} {( 4 + R )( 18 + 13 R + R^2 ) } \\
   e_9 &=& (1 + G)^2 + \frac{ (24 + 4 R) \rme^{-G} - 4 \, \rme^{-G(6 + R)/2}} {( 4 + R )( 18 + 13 R + R^2 ) } \\
\ee
Starting from state $4$, $7$ or $10$, we obtain
\be
   e_4    &=& a +        8 R\, b + R^2\, c \nonumber \\
   e_7    &=& a +  4 (R - 2)\, b - 2 R\, c \nonumber\\
   e_{10} &=& a -         16\, b +   4\, c
\ee
where
\be
   a &=& (1 + G)^2 - \frac{8}{(2 + R)^2} - \frac{21}{2 + R} + \frac{3\,( 81 + 7 R)}{18 + 13 R + R^2} \nonumber\\
   b &=& \frac{6 + R}{(2 + R)^2 (18 + 13 R + R^2)}\, \mathrm{e}^{-G(2 + R)/2} \nonumber\\
   c &=& \frac{10 + R}{(2 + R)^2 (18 + 13 R + R^2)}\, \mathrm{e}^{-G(2 + R)}
\ee
Finally, starting from state $11$ gives
\beq
   e_{11} = 1 + \frac{4 \mathrm{e}^{-2G}}{18 + 13 R + R^2} .
\eeq

} 

\newpage
\section*{References}

\bibliographystyle{prsty}

\newpage
\section*{Glossary}

\emph{Locus} %
A specific chromosomal location.
\\[1ex]
\emph{Allele} %
One of several alternative forms of a gene, or DNA sequence, at a
locus.
\\[1ex]
\emph{Genetic mosaic} %
The pattern of differences between individuals in a population.
\\[1ex]
\emph{Haplotype} %
A block of closely linked alleles that are inherited together.
Such alleles are often used as markers in the process of gene
mapping.
\\[1ex]
\emph{Linkage disequilibrium} %
At linkage equilibrium, traits at different loci are inherited
independently. Deviation from this is called linkage
disequilibrium.
\\[1ex]
\emph{Population bottleneck} %
When the population has been subject to a drastic decrease in
abundance, followed by a rapid increase in abundance. This may
happen e.g. when a small part of a population colonise a new
environment, without extensive interbreeding with the main
population.
\\[1ex]
\emph{SNP} %
Single nucleotide polymorphism. A difference in the genetic code
at a single position.
\\[1ex]
\emph{Markov process} %
A stochastic process, where the future development depends only on
the present state (no memory).
\\[1ex]
\emph{Divergence} %
When a population splits into two parts that does not interbreed,
the independent accumulation of neutral mutations within each
subpopulation leads to that the number of genetic differences
between individuals from different sub-populations increase with
time.
\\[1ex]
\emph{Gene history} %
The sequence of ancestors to a gene.
\\[1ex]
\emph{Coalescent process} %
An approximation of neutral evolution, valid for large
populations.
\\[1ex]
\emph{Chiasma process} %
Exchange of genetic material between copies chromosome pairs
during the production of gametes (egg or sperm cells).
\\[1ex]
\emph{Recombination fraction} %
The probability that two loci on the same chromosome was inherited
from different parents.

\end{document}